\documentclass[aps,article,prc,preprintnumbers,amsmath,amssymb,twocolumn,superscriptaddress,showpacs,floatfix]{revtex4}

\usepackage{epsfig}
\usepackage{dcolumn}
\usepackage{bm}

\begin{document}
\title{Universality of the Diffusion Wake from 
Stopped and Punch-Through Jets in Heavy-Ion Collisions}

\author{Barbara Betz}
\affiliation{Institut f\"ur Theoretische Physik, 
Johann Wolfgang Goethe-Universit\"at, Frankfurt am Main, Germany}
\author{Jorge Noronha}
\affiliation{Department of Physics, Columbia University, 
New York, 10027, USA}
\author{Giorgio Torrieri}
\affiliation{Frankfurt Institute for Advanced Studies (FIAS), 
Frankfurt am Main, Germany}
\author{Miklos Gyulassy}
\affiliation{Department of Physics, Columbia University, 
New York, 10027, USA}
\affiliation{Frankfurt Institute for Advanced Studies (FIAS), 
Frankfurt am Main, Germany}
\author{Igor Mishustin}
\affiliation{Frankfurt Institute for Advanced Studies (FIAS), 
Frankfurt am Main, Germany}
\author{Dirk H.\ Rischke}
\affiliation{Institut f\"ur Theoretische Physik, 
Johann Wolfgang Goethe-Universit\"at, Frankfurt am Main, Germany}
\affiliation{Frankfurt Institute for Advanced Studies (FIAS), 
Frankfurt am Main, Germany}

\begin{abstract}
We solve (3+1)--dimensional ideal hydrodynamical equations 
with source terms that describe punch-through and fully 
stopped jets in order to compare their final away-side 
angular correlations in a static medium. For fully stopped
jets, the backreaction of the medium is
described by a simple Bethe--Bloch-like model which leads to an 
explosive burst of energy and momentum (Bragg peak)
close to the end of the jet's evolution through the medium.
Surprisingly enough, we find that the medium's response 
and the corresponding away-side angular correlations are 
largely insensitive to whether the jet punches through or 
stops inside the medium.   
This result is also independent of whether momentum 
deposition is longitudinal (as generally occurs in 
pQCD energy loss models) or transverse 
(as the Bethe--Bloch formula implies).
The existence of the diffusion wake is therefore shown
to be universal to all scenarios where momentum 
as well as energy is deposited into the medium, 
which can readily be understood in ideal hydrodynamics 
through vorticity conservation.
The particle yield coming from the strong forward moving 
diffusion wake that is formed in the wake of 
both punch-through and stopped jets largely 
overwhelms their weak Mach cone signal after freeze-out.
\end{abstract}

\pacs{13.90.+i, 25.75.Bh, 25.75.Gz}
\maketitle

\section{Introduction}

One of the major discoveries found at the Relativistic 
Heavy Ion Collider (RHIC) was the suppression of highly 
energetic particles in central A+A collisions 
\cite{Adcox:2001jp,Adler:2002tq}. Jets are assumed to be 
created in the early stage of a heavy-ion collision 
where they interact with the hot and dense nuclear matter 
and serve as a hard probe for the created medium 
\cite{Gyulassy:1990ye,Gyulassy:1993hr,Wang:1994fx,Baier:1996kr,Wiedemann:2000za,Gyulassy:1999zd,Wang:2001ifa,Arnold:2001ms,Liu:2006ug,Majumder:2007zh}.
Two- and three-particle correlations of 
intermediate--$p_\bot$ particles provide an
important test of the medium response to the details of the 
jet quenching dynamics and they show a re-appearance of a 
broad or double-peaked structure in the away-side of
jet angular correlations 
\cite{Adams:2003im,Adler:2005ee,Adams:2005ph,Ulery:2005cc,Adare:2008cq,Abelev:2008nd}.

\begin{figure}[h]
\includegraphics[width=5.3cm]{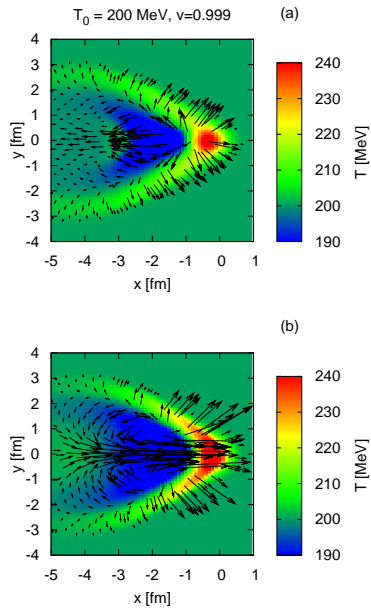}
\caption{(Color online) Temperature pattern and flow velocity 
profile (arrows) after a hydrodynamical evolution of 
$t=4.5/v_{\rm jet}$~fm, assuming (a) an energy loss rate of 
$dE/dt = 1.5$~GeV/fm for a vanishing momentum loss rate
and (b) an energy and momentum loss rate of 
$dE/dt = dM/dt = 1.5$~GeV/fm for a punch-through jet 
moving with a constant velocity of $v_{\rm{jet}}=0.999$ 
along the $x$--axis through a static background plasma 
with temperature $T_0=200$~MeV. The jet is sitting at the 
origin of the coordinates at the time of freeze-out.}
\label{fig1}
\end{figure}

The observation of strong elliptic flow in non-central 
Au+Au collisions consistent with fluid dynamical predictions 
\cite{Kolb:2003dz,Romatschke:2007mq} suggests that a 
thermalized medium that evolves hydrodynamically is 
created in these collisions. Moreover, since the average 
momentum of particles emitted on the away-side approaches the 
value of the thermalized medium with decreasing impact 
parameter \cite{Adams:2005ph}, the energy lost by the jet 
should quickly thermalize. Thus, the disturbance caused by 
the jet may also be described hydrodynamically.

Recent interest in Mach-like conical di-jet correlations 
is based on suggestions
\cite{stoeckerold1,stoeckerold2,Stoecker:2004qu,CasalderreySolana:2004qm,Satarov:2005mv,CasalderreySolana:2006sq}
that a measurement of the dependence on the cone angle 
associated with a supersonic jet moving
with velocity $v$ could provide via Mach's law 
($\cos\phi_M= c_s/v$) a constraint on the average
speed of sound in the strongly coupled Quark-Gluon Plasma 
(sQGP) \cite{Gyulassy:2004zy,Shuryak:2004cy}.
For a quantitative comparison to RHIC data, a detailed 
model of both energy and momentum deposition coupled to a
relativistic fluid model is needed \cite{CasalderreySolana:2004qm,CasalderreySolana:2005rf,CasalderreySolana:2006sq,Chaudhuri:2005vc,Renk:2006sx,Renk:2007rv,Betz:2008js,Bass:2008rv,Schenke:2008gg}.

In general, supersonic probes that shoot through a fluid 
can deposit energy and momentum in the medium in such a 
way that collective excitations such as Mach cones and diffusion
wakes are formed \cite{Landau}. These structures have 
indeed been found \cite{Chesler:2007an,Gubser:2007ga} 
in the wake of a supersonic heavy quark that travels through an 
$\mathcal{N}=4$ Supersymmetric Yang-Mills (SYM) thermal plasma 
\cite{Karch:2002sh,Herzog:2006gh,Gubser:2006bz}. 
The validity of a hydrodynamic description of the supersonic 
heavy quark wake was studied in Refs.\
\cite{Friess:2006fk,Gubser:2007ni,Chesler:2007sv,Noronha:2007xe,Gubser:2008vz}
in the framework of the Anti-de Sitter/Conformal Field Theory 
correspondence (AdS/CFT) \cite{Maldacena:1997re,Aharony:1999ti}.
The angular correlations created by heavy quark jets in 
AdS/CFT have recently been computed in Ref.\ 
\cite{Noronha:2008tg,Noronha:2008un,Gyulassy:2008fa} and 
compared \cite{Betz:2008wy} to the results obtained by 
a punch-through heavy quark jet described by the Neufeld 
{\it et. al.} \cite{Neufeld:2008fi,Neufeld:2008hs,Neufeld:2008dx} 
chromo-viscous hydrodynamic model, which is
formulated within perturbative quantum chromodynamics (pQCD).

\begin{figure}[t]
\includegraphics[scale=0.5]{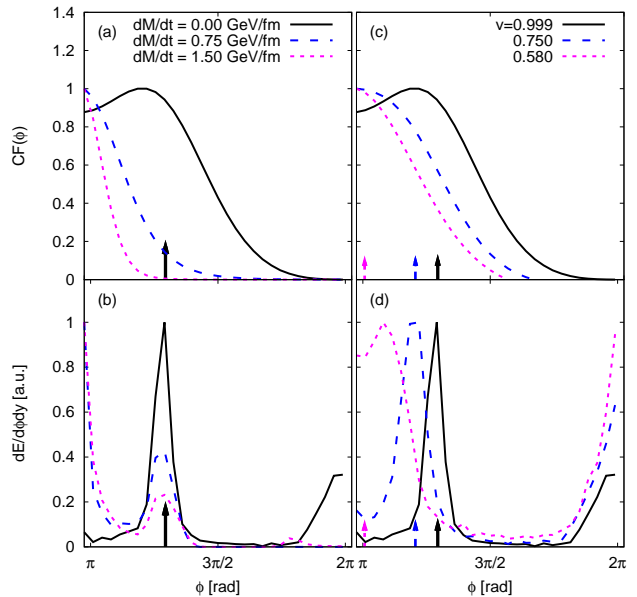}
\hspace*{0.1cm}
\caption{(Color online) The left panels show the 
normalized angular distribution created by a
punch-through jet at mid-rapidity with a fixed
energy loss of $dE/dt = 1.5$~GeV/fm and different 
momentum loss rates. The jet moves at a constant 
velocity $v_{\rm jet}=0.999$ through the medium. 
The right panels 
show the angular distributions associated with jets 
where $dE/dt = 1.5$~GeV/fm and vanishing momentum loss
($dM/dt=0$). Here, the jets move with different velocities 
through the medium: $v_{\rm jet}=0.999$ (black),
$v_{\rm jet}=0.75$ (blue), and $v_{\rm jet}=0.58$ (magenta). 
In the upper panels, an isochronous Cooper--Frye 
freeze-out at $p_\bot=5$~GeV is used while in the 
lower panels we employed the bulk flow freeze-out procedure 
\cite{Betz:2008wy}. The arrows indicate the angle of the 
Mach cone as computed via Mach's law.}
\label{fig2}
\vspace*{-0.2cm}
\end{figure}

In general, a fast moving parton (which could be a 
light quark/gluon or a heavy quark) will
lose a certain amount of its energy and momentum along its 
path through the medium and then decelerate. 
Thus, the fate of the parton jet strongly depends on its 
initial energy: if the parton has enough energy it can 
punch through the medium and fragment in the vacuum
(punch-through jet) or it can be severely quenched 
until it becomes part of the thermal bath (stopped jet). 
Of course, the amount of initial energy required for 
the parton to punch through depends on the properties of 
the medium (a very large energy loss per unit 
length $dE/dx$ means that most of the
jets will be quenched while only a few would have enough 
energy to leave the plasma). In this paper we solve the 
(3+1)--dimensional ideal hydrodynamical equations
\cite{Rischke:1995pe} with source terms that describe 
the two scenarios in order to compare the final away-side
angular correlations produced by a punch-through and a 
fully stopped jet in a static medium with
background temperature $T_0$. We would like to point out 
that the wake formed by fully stopped jets has not yet been 
studied using hydrodynamics.

For simplicity, our medium is a gas of massless $SU(3)$
gluons in which $p=e/3$, where $p$ and $e$ are the 
pressure and the energy density, respectively.
An isochronous Cooper--Frye (CF) \cite{Cooper:1974mv} 
freeze-out procedure is employed in order to
obtain the angular distribution of particles 
associated with the away-side jet. We use a
simplified Bethe--Bloch model \cite{Bethe} to show that 
the explosive burst of energy and momentum 
[known as the Bragg peak \cite{Bragg,Chen,Sihver,Kraft}]
deposited by a fully quenched jet immediately
before it thermalizes does not stop the diffusion wake 
and, thus, no new structures in the away-side of angular 
correlation functions can be found. This explosive release of
energy before complete stopping is a general 
phenomenon that has been employed, 
for instance, in applications of particle beams for 
cancer therapy \cite{WilsonPRL,Eickhoff:1999tm,pshenichnov}.

This paper is organized as follows. In Sec.\ \ref{Section2} we
describe how a jet deposits energy
and momentum in (3+1)--dimensional hydrodynamics
and how we extract observables for the Mach cone created by
the away-side jet. In Sec.\ \ref{Section3} we present our
results for punch-through, and in Sec.\ \ref{Section4} for
completely stopped jets. A summary concludes this paper
in Sec.\ \ref{Summary}.

We use natural units and the Minkowski 
metric $g_{\mu\nu}={\rm diag}(+,-,-,-)$.
Lorentz indices are denoted with Greek 
letters $\mu,\nu=0,\ldots,3$. 
In our system of coordinates,
the beam axis is aligned with the $z$--direction and
the associated jet moves along the $x$--direction 
with velocity ${\vec v}=v\,\hat{x}$.

\section{Jets in Ideal Hydrodynamics}
\label{Section2}

Energetic back-to-back jets produced in the early stages 
of a heavy-ion collision transverse to the beam axis 
can travel through the sQGP and deposit energy and momentum 
along their path in a way that depends on the physics 
behind the interaction between the jet and the underlying 
medium. In the case where one of the jets is produced 
near the surface (trigger jet), the other supersonic
away-side jet moves through the medium and excites a Mach 
wave as well as a diffusion wake. The resulting angular 
correlation with respect to the away-side jet axis is 
then expected to lead to an enhancement of associated 
hadrons at the characteristic Mach angle
\cite{Stoecker:2004qu,Satarov:2005mv,CasalderreySolana:2004qm,CasalderreySolana:2006sq,Chaudhuri:2005vc}.

In ideal hydrodynamics, the energy--momentum tensor
\begin{eqnarray}
T^{\mu\nu}= (e+p)u^\mu u^\nu - pg^{\mu\nu}\;,
\end{eqnarray}
is locally conserved, i.e.,
\begin{eqnarray}
\partial_\mu T^{\mu\nu} =0\;,
\end{eqnarray}
where $u^\mu=\gamma(1,\vec{v})$ is the 
flow 4-velocity and $\gamma=(1-\vec{v}^2)^{-1/2}$. 
We take the net baryon density to be identically 
zero in this study. Here, we only consider a 
static medium. More realistic initial conditions
involving an expanding medium will be considered in a 
further study. 

Once the jet is included in the system the conservation 
equations change. We assume that the energy lost by the jet
thermalizes and gives rise to a source 
term $S^\nu$ in the energy--momentum conservation equations
\begin{eqnarray}
\partial_\mu T^{\mu\nu} =S^\nu\;.
\label{sourceterm}
\end{eqnarray}
Thus, one has to solve Eq.\ (\ref{sourceterm}) 
numerically in order to determine the time
evolution of the medium which was disturbed by the moving jet. 
The source term that correctly depicts the interaction 
of the jet with the sQGP is unknown from first principles,
although recent calculations in AdS/CFT 
\cite{Friess:2006fk,Gubser:2007ni,Chesler:2007sv} and pQCD
\cite{Neufeld:2008hs} have shed some light on this problem. 
While pQCD is certainly the correct description in the 
hard-momentum region where jets are produced ($Q \gg T_0$), 
in the soft part of the process ($Q \sim T_0$) 
non-perturbative effects may become relevant.

In this paper, we omit the near-side correlations 
associated with the trigger jet and assume that the 
away-side jet travels through the medium according 
to a source term that depends on the jet velocity profile 
which shall be discussed below for the case
of punch-through and stopped jets.

The away-side jet is implemented in the beginning of
the hydrodynamical evolution at $x=-4.5$~fm, and
its motion is followed until it reaches $x=0$.
For a jet moving with a constant velocity $v_{\rm jet}$
this happens at $t_f = 4.5/v_{\rm jet}$~fm. 

We use two different methods to obtain the away-side 
angular correlations. In the CF method \cite{Cooper:1974mv}, 
the fluid velocity $u^\mu(t_f,\vec{x})$ and temperature
$T(t_f,\vec{x})$ fields are converted 
into free particles at a freeze-out surface
$\Sigma$ at constant time $t_f$. 
In principle, one has to ensure that energy and momentum 
are conserved during the freeze-out procedure 
\cite{CsernaiBook}. However, the associated
corrections are zero if the equation of state is 
the same before and after the freeze-out,
as it is assumed in the present study. In this case, 
the momentum distribution for associated
(massless) particles $p^\mu = \left(p_\bot,
p_\bot\cos(\pi - \phi),p_\bot\sin(\pi - \phi)\right)$
at mid-rapidity $y=0$ is computed via
\begin{equation}
\frac{dN_{\rm ass}}{p_\bot dp_\bot dy d\phi}\Big
|_{y=0}=\int_{\Sigma}d\Sigma_{\mu}p^{\mu}
\left[f_0(u^{\mu},p^{\mu},T)-f_{eq}\right]\,.
\label{cooperfrye}
\end{equation}
Here, $\phi$ is the azimuthal angle between the 
emitted particle and the trigger, $p_\bot$
is the transverse momentum, 
$f_0 = \exp[-u^\mu(t,\vec{x}) p_\mu/T(t,\vec{x})]$ the local 
Boltzmann equilibrium distribution, and 
$f_{eq}\equiv f|_{u^{\mu}=0,T=T_0}$ 
denotes the isotropic background yield.
We checked that our results do not change significantly 
if we use a Bose--Einstein distribution
instead of the Boltzmann distribution. 
The background temperature is set to $T_0=0.2$ GeV.
Following Refs.\
\cite{CasalderreySolana:2004qm,CasalderreySolana:2006sq,Betz:2008js,Noronha:2008un},
we perform an isochronous freeze-out where
$d\Sigma^\mu= d^3 \,{\vec x} \left(1,0,0,0\right)$ and 
define the angular function
\begin{equation}
CF(\phi)=\frac{1}{N_{max}}\frac{dN_{\rm ass}(\phi)}{
p_\bot dp_\bot dy d\phi}
\Big|_{y=0}\, ,
\label{CFfunction}
\end{equation}
where the constant $N_{max}$ is used to normalize the plots. 
We would like to remark that in the associated $p_\bot$--range 
of interest a coalescence/recombination hadronization scenario
\cite{Fries:2003vb,Fries:2003kq,Fries:2004hd,Greco:2003xt,Greco:2003mm}
may be more appropriate than CF freeze-out. 
However, we expect that the main features of the away-side
angular correlations obtained using CF hadronization are 
robust enough to survive other hadronization schemes.

The other freeze-out prescription 
(called bulk flow freeze-out) used in the present
paper was introduced in Ref.\ \cite{Betz:2008wy}. 
The main assumption behind the bulk flow
freeze-out is that all the particles inside a 
given small sub-volume of the fluid will be emitted 
in the same direction as the average local energy flow
\begin{equation}
\frac{d \mathcal{E}}{d\phi dy} = 
\int d^3 {\vec x}\,\, \mathcal{E}(\vec{x})\,
\delta\left[\phi - \Phi(\vec{x})\right]\, 
\delta\left[y-Y(\vec{x})\right]\,.
\label{bulkeq}
\end{equation}
Here, $\phi$ is again the azimuthal angle between the 
detected particle and the trigger jet and $y$ is the particle
rapidity. Only the $y=0$ yield is considered. 
The cells are selected according to their local azimuthal angle
$\Phi(\vec{x})=\arctan 
\left[\mathcal{P}_y(\vec{x})/\mathcal{P}_x(\vec{x})
\right]$
and rapidity
$Y(\vec{x})={\rm Artanh}\left[\mathcal{P}_z(\vec{x})/
\mathcal{E}(\vec{x})\right]$. 
The local momentum density of the cell is
$T^{0i}(\vec{x})=\mathcal{P}_{i}(\vec{x})$, 
while its local energy density in the lab frame
is $\mathcal{E}(\vec{x})=T^{00}(\vec{x})$. 
The $\delta$--functions are implemented using a
Gaussian representation as in Ref.\ \cite{Betz:2008wy}. 
Due to energy and momentum conservation,
this quantity should be conserved after freeze-out. 
Note that Eq.\ (\ref{bulkeq}) is not
restricted to a certain $p_\bot$ and does not 
include the thermal smearing that is always
present in the CF freeze-out.

\section{Punch-Through Jets}
\label{Section3}

In this section we consider a jet moving with a 
uniform velocity $v_{\rm{jet}}=0.999$
through the medium. The source term is given by
\begin{eqnarray}
\label{source}
S^\nu = \int\limits_{\tau_i}^{\tau_f}d\tau 
\frac{dM^\nu}{d\tau}\delta^{(4)}
\left[ x^\mu - x^\mu_{\rm jet}(\tau) \right],
\end{eqnarray}
where $\tau_{f}-\tau_i$ denotes the proper time interval 
associated with the jet evolution.
We further assume a constant energy and momentum loss rate
$dM^\nu/d\tau = (dE/d\tau,d\vec{M}/d\tau)$
along the trajectory of the jet 
$x^\mu_{\rm jet} (\tau) = x_0^\mu + u^\mu_{\rm jet}\tau$.
In non-covariant notation, this source term has the form
\begin{eqnarray}
\label{sourcenoncovariant}
S^\nu(t,\vec{x}) &=& \frac{1}{(\sqrt{2\pi}\,\sigma)^3}
\exp\left\{ -\frac{[\vec{x}-\vec{x}_{\rm jet}(t)]^2}{
2\sigma^2}\right\} \nonumber\\
& \times &\left(\frac{dE}{dt},\frac{dM}{dt},0,0\right)\,,
\end{eqnarray}
where $\vec{x}_{\rm jet}$ describes the location of the jet, 
$\vec{x}$ is the position on the computational grid, 
and $\sigma=0.3$. The system plasma+jet evolves 
according to Eq.\ (\ref{sourceterm}) until the freeze-out 
time $t_f=4.5/v_{\rm jet}$ fm is reached.

The temperature and flow velocity profiles 
created by a punch-through jet
with a constant energy loss rate of $dE/dt = 1.5$~GeV/fm and
vanishing momentum deposition are shown in Fig.\ \ref{fig1} (a).
In Fig.\ \ref{fig1} (b) the jet has lost the same amount 
of energy and momentum and in this case one can clearly 
see that the space-time region close to the jet, 
where the temperature disturbance is the largest, is bigger
than in the pure energy deposition scenario. The creation 
of a diffusion wake behind the jet in the
case of equal energy and momentum deposition is 
clearly visible, which is indicated by the strong
flow observed in the forward direction (at $\phi=\pi$).

Note in Fig.\ \ref{fig2} (a) that for the punch-through 
jet deposition scenario with equal energy and momentum 
loss one always obtains a peak in the associated jet 
direction after performing the freeze-out using the two 
prescriptions described in Sec.\ \ref{Section2}.
However, the energy flow distribution in Fig.\ \ref{fig2} (b) 
displays an additional small peak at the Mach cone angle 
indicated by the arrow. This Mach signal cannot be seen in the
Cooper--Frye freeze-out because of thermal smearing 
\cite{CasalderreySolana:2004qm,Noronha:2008un,Betz:2008js,Betz:2008wy}
and the strong influence of the diffusion wake,
which leads to the strong peak around $\phi\sim\pi$ 
in the bulk energy flow distribution.

However, given that the exact form of the source term in the 
sQGP is unknown, one may want to explore other 
energy--momentum deposition scenarios where the jet 
deposits more energy than momentum along its path. 
While this may seem unlikely, such a situation cannot 
be ruled out. Thus, for the sake of completeness, 
we additionally consider in Fig.\ \ref{fig2} (a) the case where
the jet source term is described by a fixed energy loss of 
$dE/dt = 1.5$~GeV/fm and different
momentum loss rates. In the bulk flow distribution in 
Fig.\ \ref{fig2} (b), one can see that
the peak at the Mach cone angle is more pronounced 
for smaller momentum loss while the contribution of the 
diffusion wake (indicated by the peak in forward direction) 
is reduced. The associated particle distribution from the CF 
freeze-out in Fig.\ \ref{fig2} (a) reveals a peak at 
$\phi \neq \pi$
for pure energy deposition (solid black line), 
however, the opening angle is shifted to a value 
smaller than the Mach cone angle 
due to thermal smearing \cite{Betz:2008js}.

In Figs.\ \ref{fig2} (c,d) we consider 
$dM/dt=0$ jets that move through the medium with
different velocities $v_{\rm jet}=0.999, 0.75$, and 
$0.58$. Note in Fig.\ \ref{fig2} (d) that the peak
position changes in the bulk flow distribution 
according to the expected Mach cone angles
(indicated by the arrows). However, due to the strong 
bow shock created by a jet moving at a
slightly supersonic velocity of $v_{\rm jet}=0.58$, 
there is a strong contribution in the forward
direction in this case and the peak position is shifted 
from the expected value. In the CF
freeze-out shown in Fig.\ \ref{fig2} (c), 
the peak from the Mach cone can again be seen for the
jet moving nearly at the speed of light ($v_{\rm jet}=0.999$), 
but for slower jets thermal smearing
again leads to a broad distribution peaked in the 
direction of the associated jet.
\begin{figure}[t]
\begin{minipage}[t]{4.2cm}
\hspace*{-0.5cm}
\includegraphics[width=4.5cm]{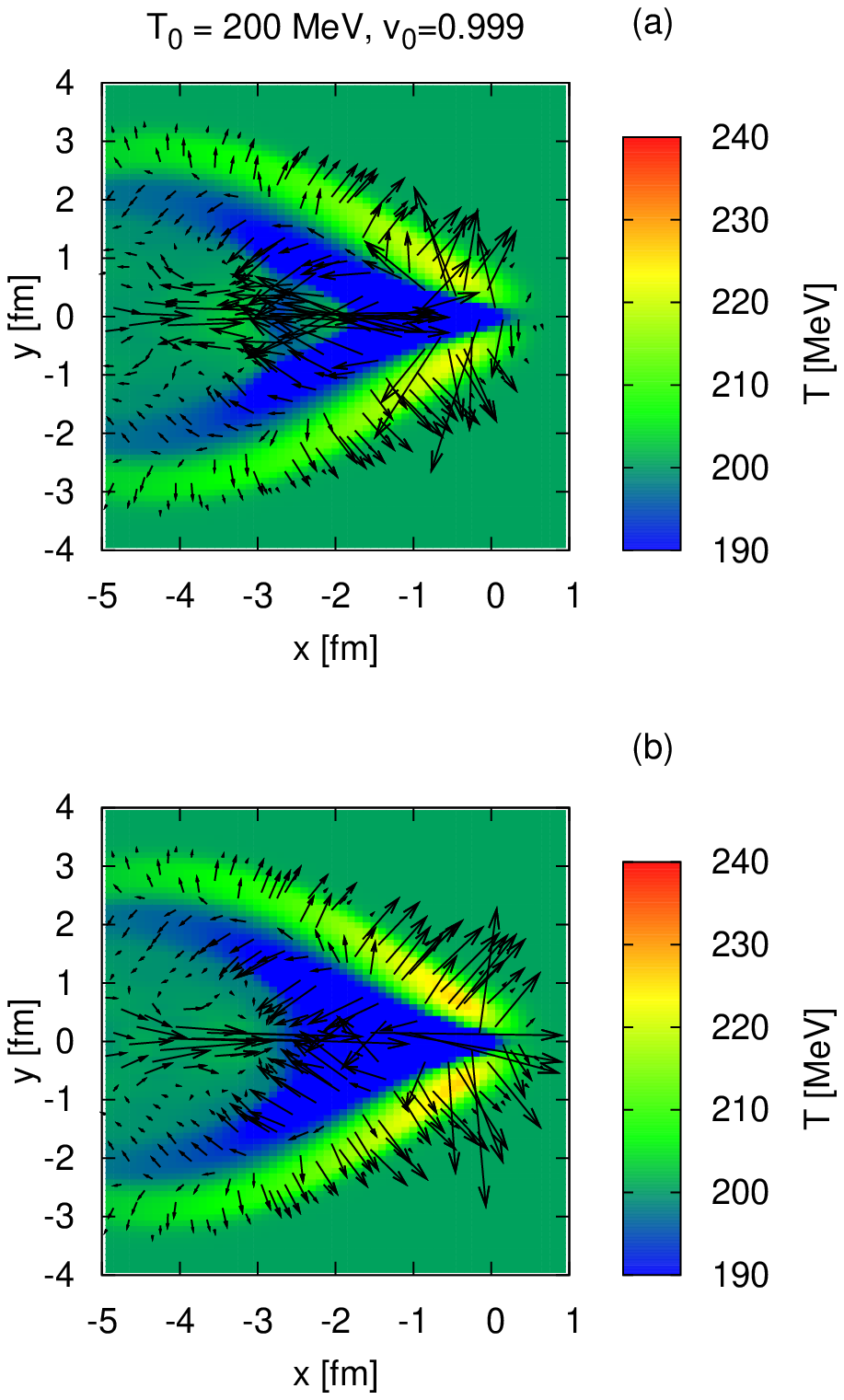}
\end{minipage}
\begin{minipage}[t]{4.2cm}
\includegraphics[width=4.5cm]{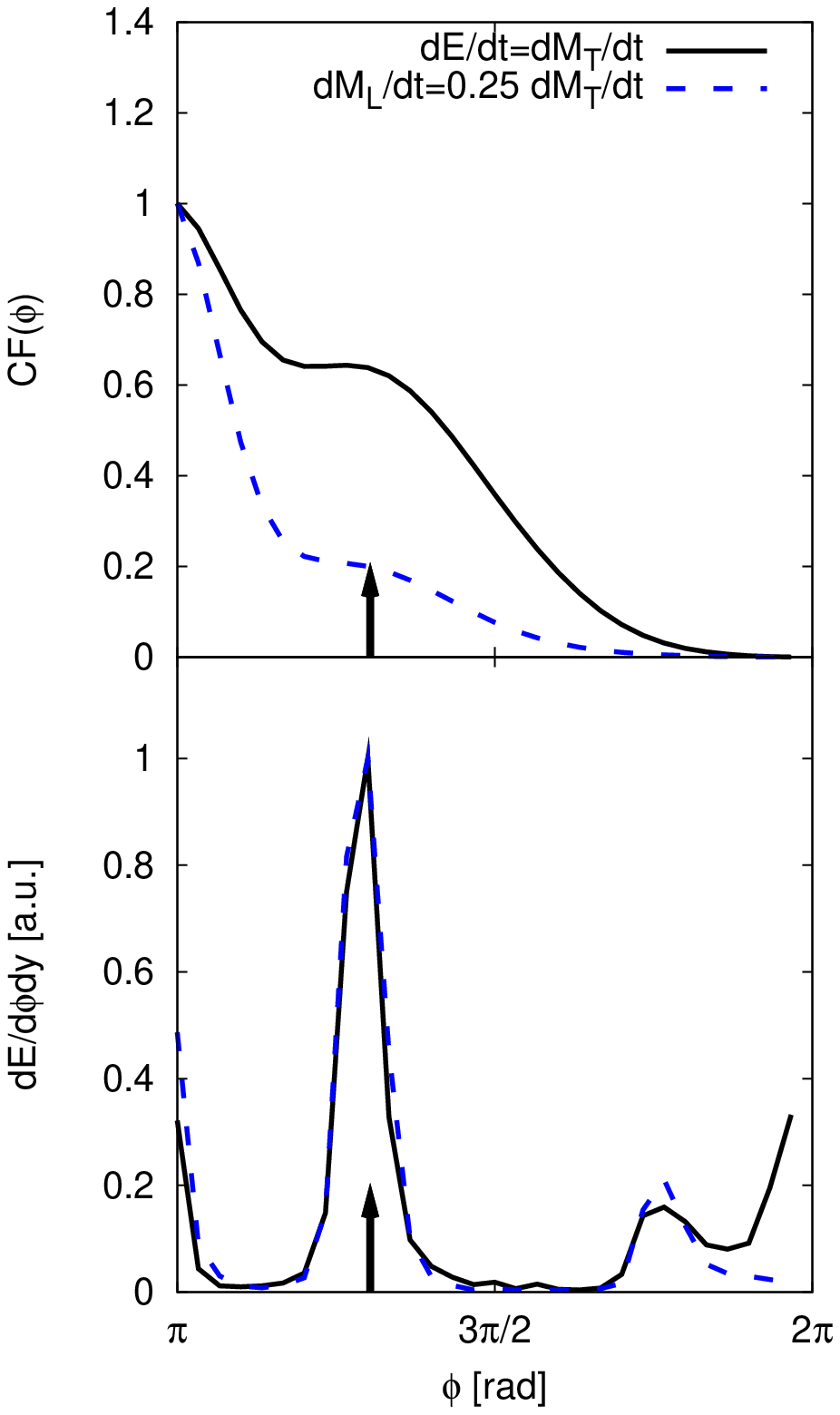}
\end{minipage}
\caption{(Color online) Left panel: Temperature pattern 
and flow velocity profile (arrows) after a 
hydrodynamical evolution of $t=4.5/v_{\rm jet}$~fm,
assuming an energy loss rate of $dE/dt = dM/dt = 1.5$~GeV/fm 
for (a) full transverse momentum deposition and 
(b) longitudinal as well as transverse momentum deposition 
with a ratio of $dM_L/dt = 0.25 \;dM_T/dt$. 
Right panel: The normalized angular distribution created by a 
punch-through jet at mid-rapidity for the two above 
mentioned transverse momentum deposition scenarios. 
In the upper panel, an isochronous Cooper--Frye freeze-out 
at $p_{\bot}=5$~GeV is used while in the lower panel the 
bulk flow freeze-out procedure  
\cite{Betz:2008wy} is employed. 
The arrows indicate the ideal Mach cone angle.}
\label{figtrans}
\end{figure}

It is apparently surprising that the above mentioned 
results are independent of whether the momentum deposited 
by the particle is in the longitudinal (along the motion of 
the jet) or transversal (perpendicular) direction.   
Repeating the calculation shown in Fig.\ \ref{fig1} 
including transverse momentum deposition
\begin{equation}
S^\nu(t,\vec{x}) \propto \left[ \begin{array}{c} dE/dt \\
dM_L/dt\\
 \left( dM_T/dt \right) \cos\varphi \\
 \left( dM_T/dt \right) \sin\varphi
 \end{array}  \right]\; ,
\label{sourcetrans}
\end{equation}
where $\varphi$ is the latitude angle 
in the $y-z$ plane with respect to the jet motion and 
the magnitude of $S^\nu(t,\vec{x})$ is the same as 
Eq.\ (\ref{source}), shows that transverse momentum 
deposition will not alter the results presented in 
this section (see Fig.\ \ref{figtrans}).
A longitudinal diffusion wake still forms during the 
fluid evolution stage, and its contribution will still 
dominate the resulting angular inter-particle correlations 
though a peak occurs around the expected Mach cone 
angle in the CF freeze-out. 

The reason is that transverse momentum deposition will 
force the fluid around the jet to expand, and the empty 
space left will create a shock wave in the longitudinal 
direction that behaves much like a diffusion wake. 
In terms of ideal hydrodynamics, this universality of the 
diffusion wake can be understood in the context of 
vorticity conservation since momentum deposition, 
whether transverse or longitudinal, will add vorticity 
to the system. This vorticity will always end up 
behaving as a diffusion wake \cite{Betz:2007kg}. 
In the next section, we demonstrate that these results are 
largely independent of whether the jet is fully quenched 
or survives as a hard trigger.

\section{Stopped Jets}
\label{Section4}

In the previous section we considered a uniformly moving 
jet that deposited energy and/or momentum in the medium 
at a constant rate. However, due to its interaction with 
the plasma, the jet will decelerate and its energy 
and/or momentum loss will change.
Thus, the deceleration roughly represents the response 
of the medium. In general, a decelerating jet should 
have a peak in the energy loss rate because the 
interaction cross section increases as the parton's 
energy decreases. In other words, when the particle's 
velocity goes to zero there appears a peak in $dE/dx$
known as the Bragg peak \cite{Bragg}.
The question to be considered in this section is whether 
this energy deposition scenario might be able to somehow 
stop the diffusion wake and, thus, change the angular 
distributions shown in Fig.\ \ref{fig2}. The source term in this
case is still given by Eq.\ (\ref{sourcenoncovariant}) and, 
according to the Bethe--Bloch
formalism \cite{Bragg,Chen,Sihver,Kraft}, one assumes that
\begin{figure}[t]
\includegraphics[scale=0.5]{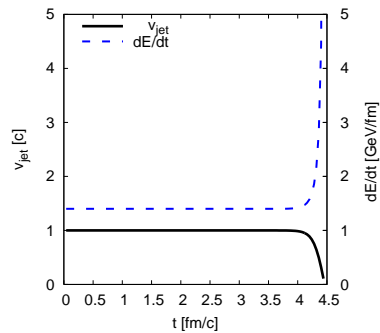}
\caption{(Color online) The jet velocity $v_{\rm jet}(t)$ 
(solid black line) and energy deposition rate
$dE(t)/dt$ (dashed blue line) according to 
Eq.\ (\ref{jetbragg}). The initial jet velocity and
energy loss rate are $v_{\rm jet}=0.999$ 
and $a\simeq-1.3607$~GeV/fm, respectively.}
\label{fig3}
\end{figure}
\begin{eqnarray}
\frac{dE(t)}{dt}=a\frac{1}{v_{\rm jet}(t)}\,,
\label{braggeqn}
\end{eqnarray}
which shows that when the jet decelerates 
the energy loss rate increases and has a peak as
$v_{\rm jet}\to 0$. Note that here $dE/dt$ is the 
energy lost by the jet, which is the negative of the
energy given to the plasma.
Using this ansatz for the 
velocity dependence of the energy loss rate and the identities
$dE/dt =v_{\rm jet}\, dM/dt$ and $dM/dy_{\rm jet}=
m\cosh y_{\rm jet}$ (as well as 
$v_{\rm jet}=\tanh y_{\rm jet}$), one can rewrite
Eq.\ (\ref{braggeqn}) as
\begin{eqnarray}
t(y_{\rm jet}) & = & \frac{m}{a}\, \left[ \sinh y_{\rm jet} 
- \sinh y_0 \right. \nonumber \\
&  & \left. - \arccos \frac{1}{\cosh y_{\rm jet}}
+ \arccos \frac{1}{\cosh y_0} \right]\hspace*{0.2ex},
\end{eqnarray}
where $y_0$ is the jet's initial rapidity. 
The equation above can be used to determine the 
time-dependent velocity $v_{\rm jet}(t)$. The initial
velocity is taken to be $v_0= {\rm Artanh} y_0 =0.999$. 
The mass of the moving parton is taken to be of the order
of the constituent quark mass $m=0.3$~GeV. Moreover, the 
initial energy loss rate $a\simeq -1.3607$~GeV/fm
is determined by imposing that the jet stops after 
$\Delta x=4.5$~fm (as in the previous section for
a jet with $v_{\rm jet}=0.999$). Thus, the jet location 
as well as the energy and momentum deposition can be
calculated as a function of time via the following equations
\begin{eqnarray}
x_{\rm jet}(t) &=& x_{\rm jet}(0) +
\frac{m}{a}\, \left[ (2-v_{\rm jet}^2)\gamma_{\rm jet} -
(2-v_{0}^2)\gamma_{0}
 \right] \;, \nonumber \\
\frac{dE}{dt} & =& a\frac{1}{v_{\rm jet}}, \hspace*{1ex}
\frac{dM}{dt} = a\frac{1}{v_{\rm jet}^2}\,,
\label{jetbragg}
\end{eqnarray}
which can be used to determine the corresponding source
term for the energy-momentum conservation equations. 
The change of the jet velocity $v_{\rm jet}(t)$ and
energy deposition $dE(t)/dt$ are displayed 
in Fig.\ \ref{fig3}. The strong increase of energy
deposition shortly before the jet is completely 
stopped corresponds to the well-known Bragg peak
\cite{Bragg}.

\begin{figure*}[t]
\includegraphics[width=17.5cm]{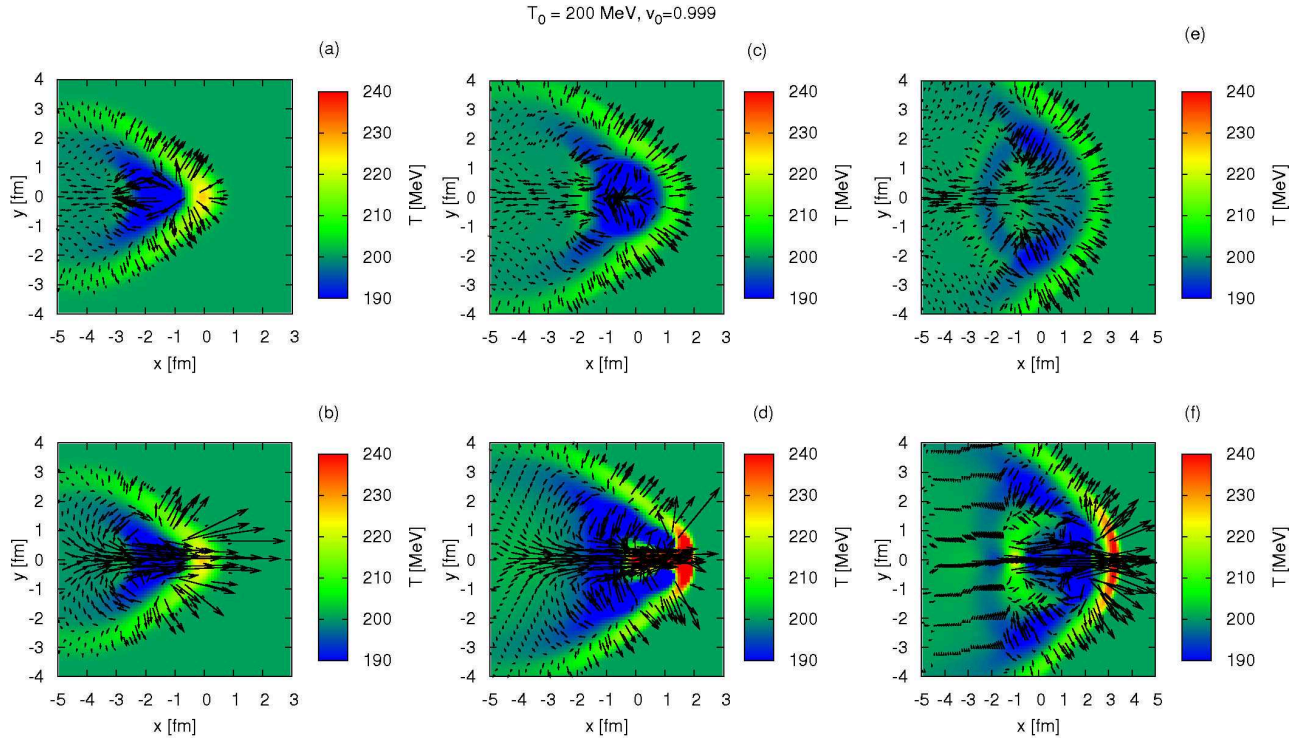}
\caption{(Color online) Temperature pattern and 
flow velocity profile (arrows) after a
hydrodynamical evolution of $t=4.5$~fm (left panel), 
$t=6.5$~fm (middle panel) and $t=8.5$~fm (right panel) 
for a jet that decelerates according to the 
Bethe--Bloch formula and stops after $\Delta x=4.5$~fm. 
The jet's initial velocity is $v_{\rm{jet}}=0.999$. 
In the upper panel a vanishing momentum loss rate is 
assumed while in the lower panel the momentum loss 
is related to the energy loss by 
Eq.\ (\ref{jetbragg}).}
\label{fig4}
\end{figure*}
The main difference between the ansatz described here 
and the Bethe--Bloch equation is that the 
momentum deposition is longitudinal 
(parallel to the motion of the jet) rather than 
transverse (perpendicular to the motion of the jet). 
According to most pQCD calculations, 
this is true in the limit of an infinite energy jet 
\cite{Gyulassy:1990ye,Gyulassy:1993hr,Wang:1994fx,Baier:1996kr,Wiedemann:2000za,Gyulassy:1999zd,Wang:2001ifa,Arnold:2001ms,Liu:2006ug,Majumder:2007zh},
but it is expected to break down in the vicinity of the 
Bragg peak where the jet energy is comparable to the energy of a
thermal particle. However, as we demonstrated in the 
previous section, the freeze-out phenomenology is rather 
insensitive to whether the momentum deposition is 
transverse or longitudinal.

Fig.\ \ref{fig4} displays the temperature and flow velocity 
profiles of a jet that stops after $\Delta x=4.5$~fm, 
with an energy loss according to Eq.\ (\ref{braggeqn}) 
and vanishing momentum deposition (upper panel) as well as an 
energy and momentum deposition following 
Eq.\ (\ref{jetbragg}) (lower panel). In the left panel the 
medium decouples immediately after $t=4.5$~fm when the 
jet is stopped while in the middle and right panel the 
decoupling takes place after $t=6.5$~fm and $t=8.5$~fm, 
respectively.

Comparing this result to Fig.\ \ref{fig1} leads to the 
conclusion that the diffusion wake is present independent
of whether the jet is quenched or survives
until freeze-out. 
In the former case, however, the diffusion wake is only
weakly sensitive to the duration of the subsequent 
evolution of the system.

Within ideal hydrodynamics this can be understood via 
vorticity conservation.  The vorticity-dominated 
diffusion wake will always be there in the ideal 
fluid, whether the source of vorticity has been quenched 
or not. The only way this vorticity can disappear 
is via viscous dissipation.  
While a (3+1)--dimensional viscous hydrodynamic calculation 
is needed to quantify the effects of this dissipation, 
linearized hydrodynamics predicts that both Mach cones and 
diffusion wakes are similarly affected \cite{Landau,CasalderreySolana:2004qm,CasalderreySolana:2006sq}.

\begin{figure*}[t]
\includegraphics[width=15.3cm]{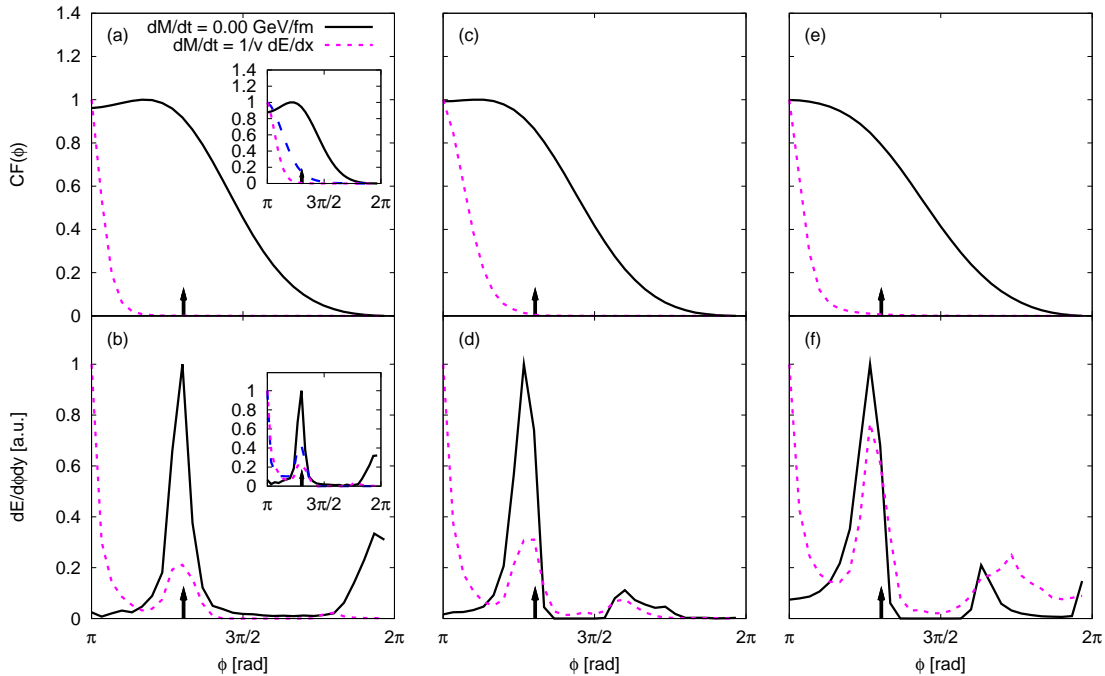}
\caption{(Color online) The normalized angular distribution 
generated by a decelerating jet (cf.\ also Fig.\ \ref{fig4}) 
at mid-rapidity is shown (upper panel) according
to an isochronous Cooper--Frye freeze-out at $p_\bot=5$~GeV 
for a jet that stops after $\Delta x=4.5$~fm and a hydrodynamical
evolution of $t=4.5$~fm (left panel), $t=6.5$~fm (middle panel) 
and $t=8.5$~fm (right panel). The corresponding bulk 
flow pattern \cite{Betz:2008wy} is shown in the lower panel. 
The solid black line in all plots depicts the pure energy
deposition case while the dashed magenta line corresponds to 
the energy and momentum deposition
scenario given by Eq.\ (\ref{jetbragg}). The arrows indicate 
the angle of the Mach cone as
computed via Mach's law. The inserts repeat 
Fig.\ \ref{fig2} (a) and (b) for comparison.}
\label{fig5}
\end{figure*}

The angular distribution associated with the decelerating jet 
(which stops after $\Delta x=4.5$~fm), shown in Fig.\ 
\ref{fig5}, is determined according to the two freeze-out 
prescriptions described in Sec.\ \ref{Section2}. When
the energy and momentum loss rates are determined by 
Eq.\ (\ref{jetbragg}) (magenta line), both freeze-out
procedures display a feature discussed in the previous 
section for the case of punch-through jets: the
formation of a strong diffusion wake which leads to a 
strong peak in the associated jet direction.
The results after the isochronous CF freeze-out 
are shown in the upper panel of Fig.\ \ref{fig5}. 
As in Fig.\ \ref{fig4}, 
the medium decouples after $t=4.5$~fm (left panel), 
$t=6.5$~fm (middle panel) and $t=8.5$~fm (right panel). 
Only the pure energy deposition scenario produces a 
peak at an angle close to the Mach angle 
[see Fig.\ \ref{fig5} (a)]
which is smeared out thermally for larger 
decoupling times [cf.\ Fig.\ \ref{fig5} (b) and (c)].
On the other hand, the bulk energy flow freeze-out 
displayed (lower panel) shows in all cases
a peak at the Mach cone angle. Note that in this case 
the peak becomes more pronounced when $dM/dt=0$. 
While the Mach cone signal increases with the decay time, 
the signal is still smaller than the forward yield of 
the diffusion wake.

\section{Summary}
\label{Summary}

In this paper we compared the away-side angular 
correlations at mid-rapidity associated with
uniformly moving jets and also decelerating jets 
in a static medium. In general, a fast moving
parton will lose a certain amount of its
energy and momentum along its path through the medium 
and thus decelerate. Therefore,
depending on its energy the jet will either punch through 
the medium and fragment in the vacuum
or it will be severely quenched until it cannot be 
distinguished from the other thermal partons in the plasma.

Our results confirm previous studies
\cite{CasalderreySolana:2004qm,CasalderreySolana:2006sq,Betz:2008js}
where a similar source term [see Eq.\ (\ref{source})] 
was used to show that the diffusion wake created by 
these jets leads to a single peak in the away-side 
of the associated di-hadron correlations that overwhelms 
the weak Mach signal after isochronous CF freeze-out
unless the total amount of momentum loss experienced 
by the jet is much smaller than the
corresponding energy loss. However, according to the 
bulk energy flow the peak always occurs at
the expected Mach cone angle but the diffusion wake still 
leads to a large peak in the associated
jet direction when $dM/dt \neq 0$ (see Fig.\ \ref{fig2}).

The same features also appear when different jet 
velocities are considered. In the bulk energy
flow distribution the peaks occur nearly at about 
the expected Mach cone angles (for slow
velocities they are shifted due to the creation of a bow 
shock), but in the CF freeze-out
distribution these peaks only occur at large jet velocities 
(see Fig.\ \ref{fig2}). This result
is consistent with the angular correlations obtained 
for a punch-through heavy quark jet
\cite{Betz:2008wy} described by the 
Neufeld {\it et. al.} pQCD source term \cite{Neufeld:2008fi,Neufeld:2008hs,Neufeld:2008dx}.

The diffusion wake created behind the jet dominates 
the freeze-out distribution for a jet moving
through a static medium, even in case of large 
transverse momentum deposition (see Fig.\ \ref{figtrans}) 
and independent of whether the jet has enough energy to 
punch through
(see Fig.\ \ref{fig1}) the medium or not (Fig.\ \ref{fig4}). 
Assuming that the jet decelerates according to the 
Bethe--Bloch formalism, see Eq.\ (\ref{braggeqn}), 
we checked whether the large amount of
energy deposited around the stopping point 
(the well-known Bragg peak) can block the diffusion wake 
and thus alter the angular correlations. 
However, our results show that no significant 
differences occur between the away-side angular 
correlations associated with punch-through jets and 
decelerating jets described within
the Bethe--Bloch model (compare Fig.\ \ref{fig2} 
and Fig.\ \ref{fig5}). Clearly, it would be interesting to 
study other models that describe decelerating jets in 
strongly-coupled plasmas. 
However, the simple Bethe--Bloch model used here 
displays the main qualitative features relevant 
for the hydrodynamic wake associated with 
decelerating jets. The path lengths of both types of 
jets were taken to be the same. A different scenario 
in which the light jets are almost 
immediately stopped in the medium while the heavy quark 
jets are still able to punch through may lead to 
different angular correlations. 
Such an analysis is left for a future study.

We would like to underline that the formation of a 
diffusion wake that trails the supersonic jet is a
generic phenomenon \cite{Landau} and, thus, its 
phenomenological consequences must be investigated 
and not simply neglected. 
Our results indicate that the diffusion wake is 
universal to all scenarios where momentum as well as 
energy is deposited to the medium,
independent of whether the jet stops or is quenched. 
However, one 
can expect that the strong forward moving column of 
fluid represented by the diffusion wake can be 
considerably distorted in an expanding 
medium by the presence of a large radial flow. The interplay 
between radial flow and away-side conical correlations in an 
expanding three-dimensional 
ideal fluid with a realistic equation of state 
\cite{NoronhaHostler:2008ju} compatible with current 
lattice results \cite{Cheng:2007jq} is currently under 
investigation.

\section*{Acknowledgments}
The authors thank I.\ Pshenichnov, L.\ Satarov, 
M.\ Bleicher, C.\ Greiner, P.\ Rau,
J.\ Reinhardt, J.\ Steinheimer and H.\ St\"ocker 
for helpful discussions and D.\ Dietrich
for carefully reading the manuscript. 
The work of B.B.\ is supported by BMBF and by the Helmholtz
Research School H-QM. J.N.\ and M.G.\ 
acknowledge support from DOE under
Grant No.\ DE-FG02-93ER40764. M.G.\ 
also acknowledges sabbatical support from DFG, ITP,
and FIAS at Goethe University. 
G.T.\ is grateful to LOEWE for the financial support given and 
thanks the Alexander von Humboldt foundation
as well as Goethe University for support. 
I.M.\ acknowledges support from the grant 
NS-3004.2008.2 (Russia).


\end{document}